\documentclass[prb,aps,showpacs,twocolumn,superscriptaddress]{revtex4}

\usepackage{graphicx}
\usepackage{amsmath}

\begin{document}

\title{Phase coherence in tight-binding models with nonrandom long-range hopping}

\author{D.\ B.\ Balagurov}
\email{d.balagurov@sns.it} \affiliation{Scuola Normale Superiore
and INFM, Piazza dei Cavalieri 7, 56126 Pisa, Italy}

\author{V.\ A.\ Malyshev}
\affiliation{S.\ I.\ Vavilov State Optical Institute, Birzhevaya
Liniya 12, 199034 Saint Petersburg, Russia} %
\affiliation{Institute for Theoretical Physics and Materials
Science Center, University of Groningen, Nijenborgh 4, 9747 AG
Groningen, The Netherlands}

\author{F. Dom\'{\i}nguez Adame}
\affiliation{GISC, Departamento de F\'{\i}sica de Materiales,
Universidad Complutense, E-28040 Madrid, Spain}


\begin{abstract}

The density of states, even for a perfectly ordered tight-binding
model, can exhibit a tail-like feature at the top of the band,
provided the hopping integral falls off in space slowly enough. We
apply the coherent potential approximation to study the
eigenstates of a tight-binding Hamiltonian with uncorrelated
diagonal disorder and long-range hopping, falling off as a power
$\mu$ of the intersite distance. For a certain interval of hopping
range exponent $\mu$, we show that the phase coherence length is
infinite for the outermost state of the tail, irrespectively of
the strength of disorder. Such anomalous feature can be explained
by the smallness of the phase-space volume for the disorder
scattering from this state. As an application of the theory, we
mention that ballistic regime can be realized for Frenkel excitons
in two-dimensional molecular aggregates, affecting to a large
extent the optical response and energy transport.

\end{abstract}

\maketitle

\section{Introduction}

The atomic orbital framework used to characterize electronic
states in solids commonly deals with short-range hopping (SRH).
For such situation it is well-known that any disorder leads to
randomization of the wave function phase on some finite spatial
scale known as phase coherence length  (PCL). Beyond this scale,
roughly equal to the mean free path, the charge transport has a
diffusive form until the coherent backscattering causes Anderson
localization.\cite{Anderson} The nature of states to be localized
or extended is governed by the dimensionality of the system.
\cite{Gang4} Extensive studies have been carried out to establish
the validity limits of the statement that all states in
one-dimensional (1D) tight-binding models are localized,
originally formulated for the case of diagonal uncorrelated
disorder and SRH.\cite{LocalizedOneD} Recently, the absence of
extended states in low-dimensional systems was questioned in
Refs.~\onlinecite{Cressoni98,Rodriguez00,MalyshevPRL,Xiong03} for
\emph{uncorrelated diagonal disorder\/} and nonrandom long-range
hopping (LRH) falling off as some power of the intersite distance.
To be specific, the authors considered the model Hamiltonian
\begin{equation}
H = \sum_{\bf n} \varepsilon_{\bf n} \left| {\bf n} \right> \!
\left< {\bf n} \right| + \sum_{{\bf n} , {\bf m}} J_{{\bf nm}}
\left| {\bf n} \right> \! \left< {\bf m} \right| \ ,
\label{Hamil}
\end{equation}
where $J_{{\bf nm}} = 1/|{\bf n} - {\bf m} |^\mu$, and
$\varepsilon_{\bf n}$ are uncorrelated random variables
distributed according to the same distribution function
$p(\varepsilon_{\bf n})$. [Energy is measured in units of the
nearest-neighbor hopping.] The size scaling of the inverse
participation ratio was investigated both numerically and with the
use of supersymmetric method for disorder averaging combined with
the renormalization group. The outcome indicated that for a
$d$-dimensional lattice ($d=1, 2$), provided $d < \mu < 3 d/2$,
the uppermost states were subjected to the Anderson
localization-delocalization transition with respect to the
disorder magnitude, remaining delocalized for not very strong
disorder. Such anomalous occurrence of extended states
dramatically differs from what has been observed so far in the
majority of localization problems.

The inverse participation ratio criterium, even though being a
robust way to detect the Anderson localization, does not
completely capture the underlying structure of the wave functions.
In particular, such criterium does not seem to distinguish the
case of pure Anderson transition from that in which the Anderson
transition is accompanied by a transition from the diffusive to
the ballistic regime. The realization of the second scenario would
imply that not only the localization length, but also the PCL
becomes infinite at the transition point. To best of our knowledge
the PCL has never been addressed so far for LRH models. At the
same time the physics of the diffusive-ballistic transition is
much simpler than that of the localization transition since it
deals essentially with the single-particle Green's function
averaged over disorder realizations.

The aim of the present paper is to investigate the single-particle
properties of the Hamiltonian~(\ref{Hamil}). In particular, we
demonstrate that in the interval of $\mu$, coinciding with that
reported in Ref.~\onlinecite{MalyshevPRL} for the existence of
extended states, the PCL diverges at the upper band-edge, even at
moderate degree of disorder. The task is fulfilled by making use
of the coherent potential approximation (CPA), known to be the
best available self-consistent approximation for the
single-particle Green's function.\cite{CPARef} The paper is
organized as follows. In Sec.~\ref{sec:MomDom} we present some
preliminary considerations of the disorder-free system, which are
necessary for a better understanding of the present paper. The
body of the paper is Sec.~\ref{sec:CPA}, where we present the CPA
approach to study the single-particle Green's function averaged
over disorder realizations. In particular, we discuss in detail
its weak-disorder asymptotic solution. Then we proceed on to
physical quantities such band edge, spectral density, density of
states (DOS) and PCL. The results of numerical simulations are
summarized in Sec.~\ref{sec:Numerical}, where we compare them to
the analytical predictions of the preceding section. We conclude
with a brief discussion of the relevance of the obtained results
in Sec.~\ref{sec:Summary}.

\section{Disorder-free system \label{sec:MomDom}}

Without disorder ($\varepsilon_{\bf n} = 0$) the eigenstates of
Hamiltonian (\ref{Hamil}) are plane waves with quasi-momenta ${\bf
k}$ within the first Brillouin zone. The corresponding
eigenenergies are given by
\begin{equation}
\label{EnergyK} E_{\bf k} = \sum_{{\bf n} \ne {\bf 0}}
\frac{e^{i{\bf k} \cdot {\bf n}}}{|{\bf n}|^\mu}\ ,
\end{equation}
where summation runs over sites of a regular $d$-dimensional
lattice ($d=1, 2$). The lattice constant is set to unity. To deal
with a bounded spectrum we assume $\mu > d$ throughout the paper.
The complete account for all terms in the sum (\ref{EnergyK}) is
important in the neighborhood of the upper band-edge, where the
LRH strongly modifies the dispersion. Specifically, around ${\bf
k} = 0$ the dispersion relation~(\ref{EnergyK}) is approximately
as follows
\begin{equation}
\label{ExpansionGeneral} E_{\bf k} = E_0 - A_d(\mu) |{\bf k}|^{\mu
- d} - B_d(\mu) |{\bf k}|^2 + O(|{\bf k}|^4)\ .
\end{equation}
Here, $E_0$, $A_d(\mu)$, $B_d(\mu)$ are known positive
constants.~\cite{Rodriguez00,MalyshevPRL} Provided $\mu < d + 2$,
the second essentially non-quadratic term in
Eq.~(\ref{ExpansionGeneral}) dominates for small $|{\bf k}|$ over
the next quadratic term, and {\it vice versa}. We therefore cast
expansion (\ref{ExpansionGeneral}) in a shorthand form
\begin{equation}
\label{ExpansionGeneral1} E_{\bf k} = E_0 - C_d(\mu) |{\bf
k}|^{\nu_d(\mu)}\ ,
\end{equation}
where
\begin{align}
\label{DefParam} C_d(\mu) & = A_d(\mu)\ , & \nu_d(\mu) & = \mu-d &
\text{for\ }  \mu < d+2\ ,
\nonumber \\
C_d(\mu) & = B_d(\mu)\ , & \nu_d(\mu) & = 2     & \text{for\ }
\mu > d+2\ .
\end{align}

Straightforward calculation of the DOS in the vicinity of the
upper band-edge yields a power-law behavior
\begin{subequations}
\begin{equation}
\label{DOSAs}
\rho(\omega) \sim a_d(\mu) |\omega - E_0|^{d/\nu_d(\mu) - 1}\ .
\end{equation}
Here, the constant factor $a_d(\mu)$ is given by
\begin{equation}
\label{DefA} a_d(\mu) = \frac{S_d}{(2\pi)^d} \,
\frac{[C_d(\mu)]^{-d/\nu_d(\mu)}}{\nu_d(\mu)}\ ,
\end{equation}
\end{subequations}
with $S_d$ being the area of the $d$-dimensional unit sphere (we
set $S_1 = 2$). It should be noticed that the DOS~(\ref{DOSAs}) is
very sensitive to the value of $\mu$: the exponent $d/\nu_d(\mu) -
1$ in~(\ref{DOSAs}) has the familiar Van Hove form $(d-2)/2$ for
$\mu > d + 2$ and the less usual one, involving the dependence on
$\mu$, for the opposite inequality. Furthermore, as $\mu < 3d/2$,
both the DOS and its derivative vanish at the energy $E_0$,
indicating that this part of the energy spectrum is weakly
populated by the states. In spite of a qualitative resemblance of
a disorder-induced band tail, in the model under consideration
this is a purely kinetic feature, stemming from the long-range
nature of hopping.

Besides the unusual form of the DOS, the LRH results not in
exponential but rather a power-law localization of states.
Previously found in numerical simulations,\cite{PLLoc} this
feature can be easily understood considering a nontypical site
energy fluctuation, i.e. the one with site energy
$\varepsilon_{\bf n}$ essentially outside of the band. Then, the
first order of perturbation theory with respect to the
off-diagonal part of the Hamiltonian~(\ref{Hamil}) yields the
corresponding wave function $\Phi_{{\bf nm}} = \delta_{\bf nm} +
J_{\bf nm}/\varepsilon_{\bf n}$ which falls off as $1/|{\bf
n-m}|^\mu$ upon increasing the distance from the site ${\bf n}$.
Remarkably, $\Phi_{{\bf nm}}$ does not vanish outside some finite
real-space interval, as it would be for a perturbative wave
function in the case of SRH. This means that for the LRH the
perturbation theory provides an approximate solution valid in all
space, and no higher-order corrections are needed as long as
$|\varepsilon_{\bf n}| \gg |J_{\bf nm}|$. For a generic relation
between $\varepsilon_{\bf n}$ and $J_{\bf nm}$ the omitted orders
will correct the coefficient of the $1/|{\bf n-m}|^\mu$ dependence
and introduce some new terms falling off faster than $1/|{\bf
n-m}|^\mu$. For $\varepsilon_{\bf n}$ inside the band, both
power-law and exponentially falling-off components in the wave
function will appear. Their form can be deduced from calculations
presented in Appendix~\ref{app:oned} for the case of 1D system.

\section{Coherent-potential approximation \label{sec:CPA}}

The CPA is a reliable and efficient method to study the
disorder-averaged DOS and the phase coherence between wave
functions for different members of the random
ensemble.\cite{CPARef,EliottRev} These characteristics can be
extracted from the disorder-averaged single-particle propagator
\begin{equation}
\label{GDef} G(\omega) = \int \prod_{\bf n} d \varepsilon_{\bf n}
p(\varepsilon_{\bf n}) \frac1{\omega - H} \ .
\end{equation}
Unless otherwise specified, the energy variable $\omega$ will have
an infinitesimal positive imaginary part. In the framework of CPA,
quantity~(\ref{GDef}) is approximated in a self-consistent way to
preserve its analytic properties and deliver the correct limiting
behavior as either the disorder strength or the hopping amplitude
tend to zero. Namely, it is argued that a good approximation for
the single-particle propagator is to treat its disorder-generated
self-energy as a site-diagonal quantity. In other words,
average~(\ref{GDef}) is evaluated by replacing the site energy
$\varepsilon_{\bf n}$ by a coherent potential $\sigma(\omega)$
\begin{equation}
\label{GreenK}
G_{\bf k}(\omega) = \frac1{\omega - \sigma(\omega) - E_{\bf k}}\ .
\end{equation}
The choice of $\sigma(\omega)$ should be such to compensate on
average scattering from a single site. The resulting
self-consistency condition reads (see, e.g.,
Ref.~\onlinecite{EliottRev})
\begin{equation}
\label{CPASelfCons} \int d \varepsilon_{\bf n} p(\varepsilon_{\bf n})
\frac{\varepsilon_{\bf n} - \sigma(\omega)}{1 - [\varepsilon_{\bf n} -
\sigma(\omega)] G_{\bf nn}(\omega)} = 0\ ,
\end{equation}
where $G_{\bf nn}(\omega)$ is the site-diagonal element of the CPA
Green's function.

We adopt for the random site energies the distribution uniform
within symmetric interval $[-\Delta, \Delta]$. For such
$p(\varepsilon_{\bf n})$, the integral in Eq.~(\ref{CPASelfCons})
can be evaluated explicitly, and we get the self-consistency
equation in the form
\begin{equation}
\label{CPALog}
\frac1{2\Delta} \ln \frac{1 + [\sigma(\omega) +
\Delta] G_{\bf nn}(\omega)}{1 + [\sigma(\omega) - \Delta] G_{\bf
nn}(\omega)} = G_{\bf nn}(\omega)\ .
\end{equation}
The resulting CPA propagator will be studied numerically in
Sec.~\ref{sec:Numerical}.  In the rest of the present section, we
derive an analytic solution of the theory in the weak-disorder
limit. As an outcome, the asymptotic formulas for the observable
quantities, namely band-edge energy, spectral density, DOS and PCL
will be obtained.

\subsection{Weak-disorder self-energy \label{sec:LowDisSE}}

To access the weak-disorder limit of CPA we perform a formal
small-$\Delta$ expansion in Eq.~(\ref{CPALog}) retaining terms up
to $\Delta^2$. Then the self-consistency condition acquires the
form
\begin{equation}
\label{MyEq1}
\sigma(\omega) = \frac{\Delta^2}3 \,
\frac{G_{\bf nn}(\omega)}{[1 + G_{\bf nn}(\omega) \sigma(\omega)]^2} \ .
\end{equation}
The term $G_{\bf nn}(\omega) \sigma(\omega)$ in the denominator
can be neglected because eventually it turns out to be
proportional to a positive power of $\Delta \ll 1$. Thus the
weak-disorder equation to be analyzed reads
\begin{equation}
\label{SelfConsReduced} \sigma(\omega) = \frac{\Delta^2}{3} \,
G_{\bf nn}(\omega).
\end{equation}
Both of the two approximative steps yielding
Eq.~(\ref{SelfConsReduced}) do not introduce any spurious
singularities around the band edges, e.g.~like those emerging
after truncation of the cumulant series for the Green's function
(see Ref~\onlinecite{EliottRev}). Such preservation of the
analyticity in the vicinity of the band edge is guaranteed by the
full account for self-energy in the local Green's function
entering the equation. We notice that the coefficient
$(1/3)\Delta^2$ in Eq.~(\ref{SelfConsReduced}) coincides with the
site-energy variance, the second centered moment of
$p(\varepsilon_{\bf n})$. From this remark, we conclude that the
weak-disorder equation~(\ref{SelfConsReduced}) holds for any
distribution function $p(\varepsilon_{\bf n})$ decreasing fast
enough at large $\varepsilon_{\bf n}$ (specifically, if the
$M$th-order centered moment scales as $\Delta^M$).

Weak disorder mixes only a small part of otherwise
well-defined-momenta states within an energy interval of the order
of the effective broadening $\sim \,{\rm Im}\, \sigma(\omega)$.
This facilitates evaluation of the site-diagonal propagator
entering Eq.~(\ref{SelfConsReduced}). Namely, for energies close
to the upper band edge only small $\bf k$ will contribute to
$G_{\bf nn}(\omega) = \sum_{\bf k} G_{\bf k}(\omega)$. Because
$\sigma(\omega)$ is small due to the weakness of disorder, while
the detuning $\omega - E_0$ can be made small close enough to the
upper band edge, the site-diagonal Green's function can be found
as
\begin{equation}
\label{GSmallK} G_{\bf nn}(\omega) \sim \frac{\pi
a_d(\mu)\big[\omega - E_0 - \sigma(\omega)\big]^{d/\nu_d(\mu)-1}}
{\sin [\pi d / \nu_d(\mu)]} + b_d(\mu).
\end{equation}
Here, the first term [in which $a_d(\mu)$ is given by
Eq.~(\ref{DefA})] comes solely from the small momenta and does not
depend on any momentum cut-off. The magnitude of the contributing
$\bf k$ can be estimated from the dispersion law
(\ref{ExpansionGeneral1}) as $k \sim |\omega - E_0 -
\sigma(\omega)|^{1/\nu_d(\mu)}$. The imaginary part of the
considered term is nonzero in the limit of vanishing disorder
[$\sigma(\omega)=0$], in agreement with the singular or tail-like
DOS of Eq.~(\ref{DOSAs}). The second term in Eq.~(\ref{GSmallK})
is a {\it real} constant representing the part of the
site-diagonal propagator whose dependence on energy is smooth near
the upper band edge; the major contribution comes from the higher
$\bf k$. The exact value of $b_d(\mu)$ can be found via numerical
summation over all momenta in $G_{\bf nn}(\omega)$.

Accounting for the constant $b_d(\mu)$ in the weak-disorder
self-consistency equation (\ref{SelfConsReduced}) is equivalent to
a shift of both the band edge position and the self-energy.
Namely, upon transformation $E'_0 = E_0 + (1/3) \Delta^2
b_d(\mu)$, $\sigma'(\omega) = \sigma(\omega) - (1/3) \Delta^2
b_d(\mu)$, Eq.~(\ref{SelfConsReduced}) takes a form as if only the
low-momenta states were participating in the disorder scattering:
\begin{equation}
\label{SelfConsRen} \sigma'(\omega) = \frac{\Delta^2}{3}\,
\frac{\pi a_d(\mu)\big[\omega - E'_0 - \sigma'(\omega)
\big]^{d/\nu_d(\mu)-1}} {\sin[\pi d / \nu_d(\mu)]} \ .
\end{equation}
While $E'_0 - E_0$ is a mere shift the uppermost states acquire
due to the coupling to the continuum of remote low-energy
(high-momenta) states, the broadening effect, accounted through
Eq.~(\ref{SelfConsRen}), originates from the mixing of nearby
states at the band edge.

The solution of Eq.~(\ref{SelfConsRen}) turns out to be extremely
sensitive to the value of the exponent $\mu$. In particular, for a
fixed energy $\omega = E'_0$ it follows that either
\begin{subequations}
\begin{equation}
\label{nu>d/2}
    \sigma'(E'_0) = \left\{ \frac{\Delta^2}{3}
    \frac{\pi a_d(\mu)}{\sin[\pi d/\nu_d(\mu)]}
    \right\}^{\nu_d(\mu) / [2 \nu_d(\mu) - d]}
\end{equation}
or
\begin{equation}
\label{nu<d/2}
    \sigma'(E'_0) = 0 \ .
\end{equation}
\end{subequations}
The former expression can be considered as being valid only at
$\mu > 3d/2$ where $\nu_d(\mu) > d/2$, because otherwise (at $\mu
< 3d/2$) the exponent $\nu_d(\mu) / [2\nu_d(\mu) - d]$ becomes
negative, resulting in failure of the zero-disorder limit. Then,
the latter expression should be used.

From the presented arguments it follows that the renormalized
self-energy remains finite at $\omega = E'_0$ varying with energy
on the scale $\sim \Delta^{2\nu_d(\mu)/[2\nu_d(\mu) - d]}$
provided $\mu > 3d/2$. The complete solution of
Eq.~(\ref{SelfConsRen}) in this case looks rather involved, but
for the forthcoming, qualitative considerations based on disorder
scaling
\begin{subequations}
\begin{equation}
\label{SigmaScaling1}
\sigma'(\omega) \sim \Delta^{2\nu_d(\mu)/[2\nu_d(\mu) - d]}
\end{equation}
will suffice. Regarding the range of $\mu < 3d/2$, the
renormalized self-energy vanishes faster than linearly as $\omega$
approaches $E'_0$. The solution in this case can be obtained in a
closed form
\begin{equation}
\label{SigmaScaling2}
    \sigma'(\omega)\sim\frac{\Delta^2}3\frac{\pi a_d(\mu)}
    {\sin[\pi d/\nu_d(\mu)]}(\omega -E'_0)^{d/\nu_d(\mu)-1}\ .
\end{equation}
\end{subequations}

Having derived the CPA self-energy, at next step we present
analytical results concerning observable quantities, namely
band-edge energy, spectral density, DOS, and PCL.

\subsection{Band-edge energy \label{sec:BEE}}

The CPA band edge, denoted by $\tilde E$, is determined as the
maximum energy at which $\,{\rm Im}\,\sigma(\omega) \ne 0$. As
follows from the above considerations, $\tilde E$ is shifted from
its disorder-free value $E_0$, first, by $(1/3)\Delta^2 b_d(\mu)$
due to scattering processes involving large $\bf k$ states, and
second, by an amount $\sim \Delta^{2\nu_d(\mu) /[2\nu_d(\mu) -
d]}$ giving the scale on which the renormalized self-energy varies
significantly. Comparing the disorder scaling exponent appearing
in the two contributions, we conclude that the former prevails
over the latter provided $\nu_d(\mu) > d$. Since it is always
$\nu_d(\mu) \le 2$ in 2D systems, the disorder-induced shift
$\tilde E - E_0$ scales as $\Delta^2$. In 1D system the scaling
depends on the hopping exponent: for $\mu < 2$ one gets $\tilde E
- E_0 \sim \Delta^2$, while $\tilde E - E_0 \sim
\Delta^{2(\mu-1)/(2 \mu - 3)}$ for $\mu>2$.

\subsection{Spectral density}

Along with the results reported in Sec.~\ref{sec:LowDisSE}, one
can also obtain the disorder-averaged spectral density at the
upper band-edge:
\begin{equation}
\label{A}
    A_{\bf k}(\omega) = -\frac{1}{\pi} \,{\rm Im}\,G_{\bf k}(\omega)\ .
\end{equation}
From Eq.~(\ref{GreenK}) the energy-domain width of the spectral
density can be estimated as $\gamma \sim |\,{\rm
Im}\,\sigma(\omega)|$. As has been demonstrated above, if the LRH
exponent $\mu$ is greater than $3d/2$, such width will scale upon
disorder as
\begin{equation}
\label{GammaScal} \gamma \sim \Delta^{2\nu_d(\mu) /[2\nu_d(\mu)
-d]} \ .
\end{equation}
This formula is a generalization of the disorder-induced linewidth
estimate $\gamma \sim \Delta^{4/3}$ known for 1D tight-binding
models with SRH. \cite{LitScaling,Malyshev91} Upon increasing the
hopping range (decreasing $\mu$) the disorder broadening of the
resonance becomes less pronounced. As $\mu$ goes below $3d/2$, the
zero-momentum spectral density acquires a power-law form
\begin{equation}
\label{Absorption}
A_{{\bf k}=0}(\omega) \sim \Delta^2 |\omega -E'_0|^{d/\nu_d(\mu)-3}\ .
\end{equation}
Noticeably, for the specified interval of $\mu$ the exponent
$d/\nu_d(\mu) - 3$ is greater than $-1$. This guarantees
integrability of the spectral density around $\omega = E'_0$.
Hence, there is no contradiction with the single-particle sum
rule.

\subsection{Density of states}

The next important question we address is the asymptotic behavior
of the DOS in the proximity of the band edge.
Expression~(\ref{GSmallK}) can be used to relate the DOS
\begin{equation}
\label{rho} \rho(\omega) = -\frac{1}{\pi} \,{\rm Im}\, G_{\bf
nn}(\omega) \,
\end{equation}
to the CPA self-energy. As follows from
Eq.~(\ref{SelfConsReduced}), the DOS drops abruptly, i.e.~has
infinite derivative at the band-edge, provided $\mu > 3d/2$. In
the opposite case, $\mu < 3d/2$, using Eq.~(\ref{SigmaScaling2})
we obtain
\begin{equation}
\label{DOSApp}
\rho(\omega) \sim a_d(\mu) |\omega -E'_0|^{d/\nu_d(\mu)-1}\ .
\end{equation}
Here, the DOS profile remains similar to that of the
non-disordered system [Eq.~(\ref{DOSAs})], the only difference
being the band-edge location.

\subsection{Phase coherence length\label{sec:PCL}}

The PCL, to be denoted by $N(\omega)$, is usually defined as the
inverse exponent responsible for exponential fall-off of the
coordinate-space single-particle propagator.\cite{Abrikosov} Such
definition cannot be straightforwardly adopted in the case of LRH
because, as demonstrated in Appendix~\ref{app:oned}, the
propagator falls off in essentially non-exponential way. Instead,
we shall relate the PCL to an appropriately measured
momentum-domain width of the spectral function $\kappa(\omega)$,
through $N(\omega) = 1/\kappa(\omega)$. For energies close to the
upper band edge, $\kappa(\omega)$ is estimated from
Eqs.~(\ref{ExpansionGeneral1}) and~(\ref{GreenK}) to be
\begin{subequations}
\begin{equation}
\label{KappaDef}
\kappa(\omega) \sim \,{\rm Im}\, \left[ -
\frac{\omega - E_0 - \sigma(\omega)}{C_d(\mu)} \right]^{1/\nu_d(\mu)}\ .
\end{equation}
Using Eq.~(\ref{SigmaScaling1}) for the CPA self-energy, one gets
the following disorder scaling of the PCL valid in the range $\mu
> 3d/2$:
\begin{equation}
\label{NScal1} N(\omega) \sim \Delta^{-2/[2\nu_d(\mu)-d]}\ .
\end{equation}

It is worth to notice that scaling~(\ref{NScal1}) is nicely
reproduced by means of a simple argument similar to that used in
Ref.~\onlinecite{Malyshev91} for the SRH model. One proceeds
confronting the two quantities:~\cite{Malyshev91} energy spacing
$\delta E$ between two adjacent states localized within domain of
a linear size $N$, and the reduced disorder magnitude $\sim \Delta
N^{-d/2}$ seen by the localized quasiparticle, the effect known as
exchange narrowing.~\cite{Knapp84} Using value $N$ as a
quantization length, from Eq.~(\ref{ExpansionGeneral1}) we can
estimate the energy spacing $\delta E \sim N^{-\nu_d(\mu)}$.
Applying condition $\delta E \sim \Delta N^{-d/2}$, one arrives at
the disorder scaling of $N$ as is given in Eq.~(\ref{NScal1}).
Furthermore, the spectral density width~(\ref{GammaScal}) is
recovered being identified with the exchange-narrowed disorder
magnitude $\sim \Delta N^{-d/2}$ where $N$ is found above.

For $\mu < 3d/2$, special attention should be paid on the fact
that $\sigma'(\omega)$ vanishes faster than linearly as $\omega$
approaches $E'_0$. Direct calculation shows that in this case, the
PCL inside the spectral region~($\omega < E'_0$) scales as
\begin{equation}
\label{NScal2}
N(\omega) \sim \frac1{\Delta^2} |\omega - E'_0|^{2 - (d+1)/\nu_d(\mu)}\ .
\end{equation}
\end{subequations}
Remarkably, such $N(\omega)$ diverges not only upon decreasing the
strength of disorder $\Delta$, but also as energy $\omega$
approaches the band edge. From both Eqs.~(\ref{NScal1})
and~(\ref{NScal2}) it follows that the PCL is infinite at the
upper band edge in the marginal case $\mu=3d/2$.

\section{Numerical results \label{sec:Numerical}}

\begin{figure}[ht]
\centering
\includegraphics[width=85mm]{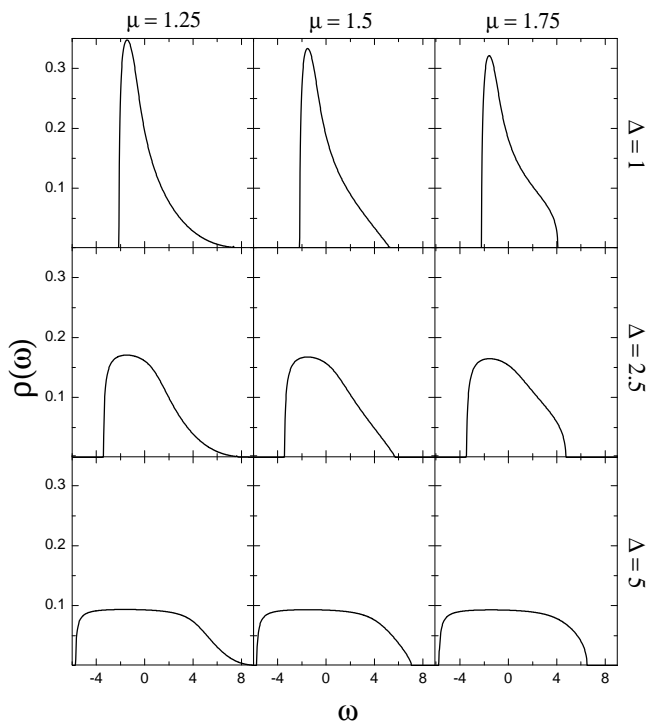}\\[3mm]
\caption{CPA DOS calculated for several values of disorder
strength $\Delta$ and hopping-range exponents $\mu$.}
\label{fig:Rho1}
\end{figure}

To further clarify the properties possessed by the LRH
model~(\ref{Hamil}), we solved the CPA equations numerically,
without any of approximations used in the above analytical
treatment. In this section, we present the numerical results and
compare them to those obtained in Sec.~\ref{sec:CPA}. The
self-consistency equation~(\ref{CPALog}) was solved using a
standard iterative scheme; the momentum integral~$G_{\bf
nn}(\omega) = \sum_{\bf k} G_{\bf k}(\omega)$ was evaluated with
an efficient mesh-optimized algorithm. For definiteness, we
restricted ourselves to one dimension, $d=1$. To check the
accuracy of CPA, we also directly diagonalized the
Hamiltonian~(\ref{Hamil}) for an open chain of ${\cal N}=1000$
sites with statistical averaging over $500$ disorder realizations.

\begin{figure}[ht]
\centering
\includegraphics[width=70mm]{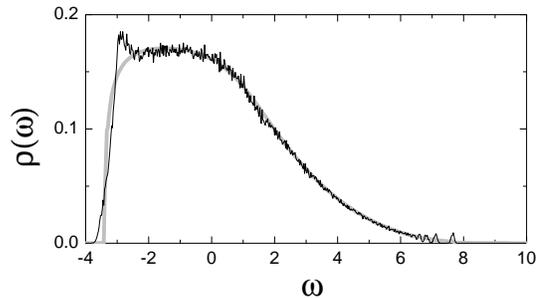}
\caption{CPA DOS (bold line) compared to that of exact
diagonalization (narrow line) for $\mu = 1.25$ and $\Delta=2.5$.}
\label{fig:Rho2}
\end{figure}

The calculated DOS profiles are presented in Fig.~\ref{fig:Rho1}
for $\mu=1.25$, $\mu=1.5$ (marginal case), and $\mu=1.75$, and
several values of the disorder strength. We notice difference in
the behavior of the CPA DOS at the high-energy tail depending on
the exponent $\mu$, in accordance with the previous discussion. We
also observe that the low-energy side of the DOS is almost
independent of $\mu$ since the dispersion relation of the
disorder-free system is parabolic at the bottom of the
band.\cite{MalyshevPRL} As illustrated in Fig.~\ref{fig:Rho2},
even for noticeable disorder ($\Delta = 2.5$) the CPA is in
excellent agreement with the exact diagonalization. Such
impressive accuracy of CPA for tight-binding models with simple
band structure is guaranteed by the high number of energy-domain
moments of the spectral density being reproduced
exactly.\cite{Moments} At the same time, the fully numerical
spectrum suffers from finite-size oscillations, especially
noticeable in the DOS tail for values of $\mu$ close to unity. The
lack of smoothness is an unavoidable consequence of the smallness
of DOS in this spectral region. To estimate the finite-size
contribution one can look at spacing $\delta E$ between the ground
and the first size-quantization levels with the disorder being
turned off. As follows from Eq.~(\ref{ExpansionGeneral1}), such
spacing scales with length $\cal N$ of the chain as $\delta E \sim
{\cal N}^{1- \mu}$. The hopping integral falling-off slowly enough
($\mu - 1 \ll 1$) leads to the rapidly growing size of matrices
needed to avoid the discreteness of DOS in the tail region. For
instance, to get a reasonably small value $\delta E = 0.01$ for
$\mu = 1.25$, one would have to handle numerically the matrices of
rather generic structure with size ${\cal N} \sim 10^8$, the task
unaffordable for computers.

\begin{figure}[ht]
\centering
\includegraphics[width=60mm]{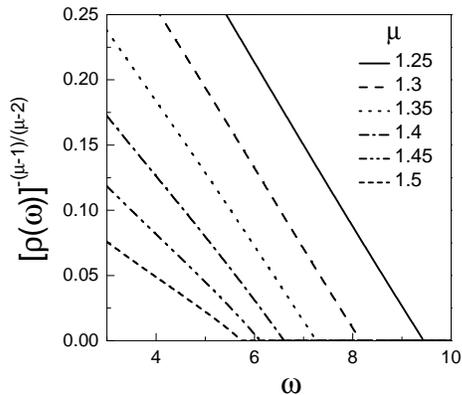}
\caption{CPA DOS after the power transformation. The degree of
disorder $\Delta$ is $2.5$ for all curves.} \label{fig:PowerRho}
\end{figure}

\begin{figure}[ht]
\centering
\includegraphics[width=70mm]{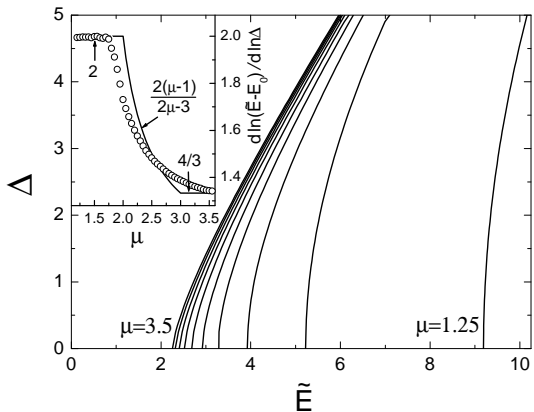}
\caption{Dependence of the CPA upper band-edge on the degree of
disorder. Curves are plotted for several fixed $\mu$ selected
between $1.25$ and $3.5$ with step $0.25$. The open circles in the
inset show the exponent obtained after fitting $\tilde E - E_0$
with a power-law function of disorder. The same exponent predicted
by the weak-disorder analytic solution is shown with solid line.}
\label{fig:BandEdge}
\end{figure}

To check the analytic expression~(\ref{DOSApp}) against the full
CPA solution we plotted in Fig.~\ref{fig:PowerRho} a power of the
CPA DOS with the exponent equal to the inverse of that in
formula~(\ref{DOSApp}) with $d=1$. The fact that all curves with
$\mu$ not exceeding $3/2$ can be fitted to a straight line near
the upper band edge confirms the validity of the
asymptotic~(\ref{DOSApp}). The shift of the upper band edge with
respect to its disorder-free location can be extracted from the
data presented in Fig.~\ref{fig:PowerRho}. The CPA results
concerning the band edge are summarized in
Fig.~\ref{fig:BandEdge}. The figure illustrates the general
tendency of the band edge to become more sensitive to disorder
with increasing the range of hopping. We also verified the scaling
arguments of Sec.~\ref{sec:BEE} concerning the band edge by
fitting the displacement $\tilde E - E_0$ to a power-law of
$\Delta$. The outcome, summarized in the inset of
Fig.~\ref{fig:BandEdge}, demonstrates a qualitative agreement
between the weak-disorder analytical considerations and the full
CPA solution. The agreement is especially good in the extreme
limits of very large and very small range of hopping ($\mu - 1 \ll
1$ and $\mu > 3$, respectively). The discrepancy in the
intermediate region is caused by an insufficient precision while
separating out the dominating power-law contribution in the
presence of other similar contributions with close exponents.

\begin{figure}[ht]
\centering
\includegraphics[width=50mm]{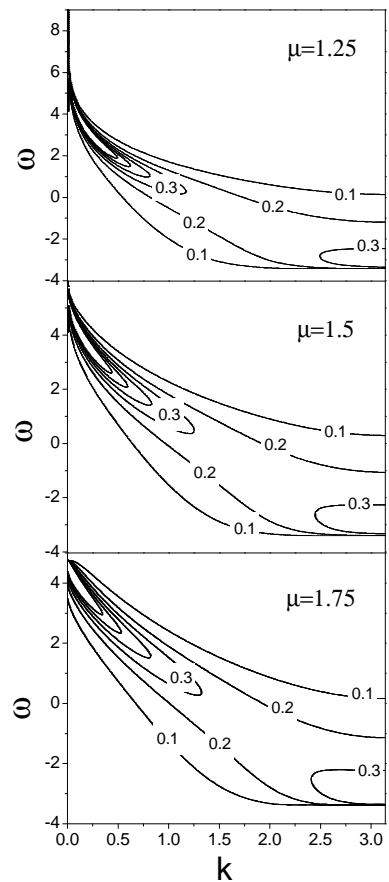}
\caption{Contour plots of the CPA spectral density $A_{\bf
k}(\omega)$ for $\mu=1.25$, $1.5$, $1.75$ and degree of disorder
$\Delta=2.5$.} \label{fig:SDMomentum}
\end{figure}

\begin{figure*}[t]
\centering
\includegraphics[height=85mm]{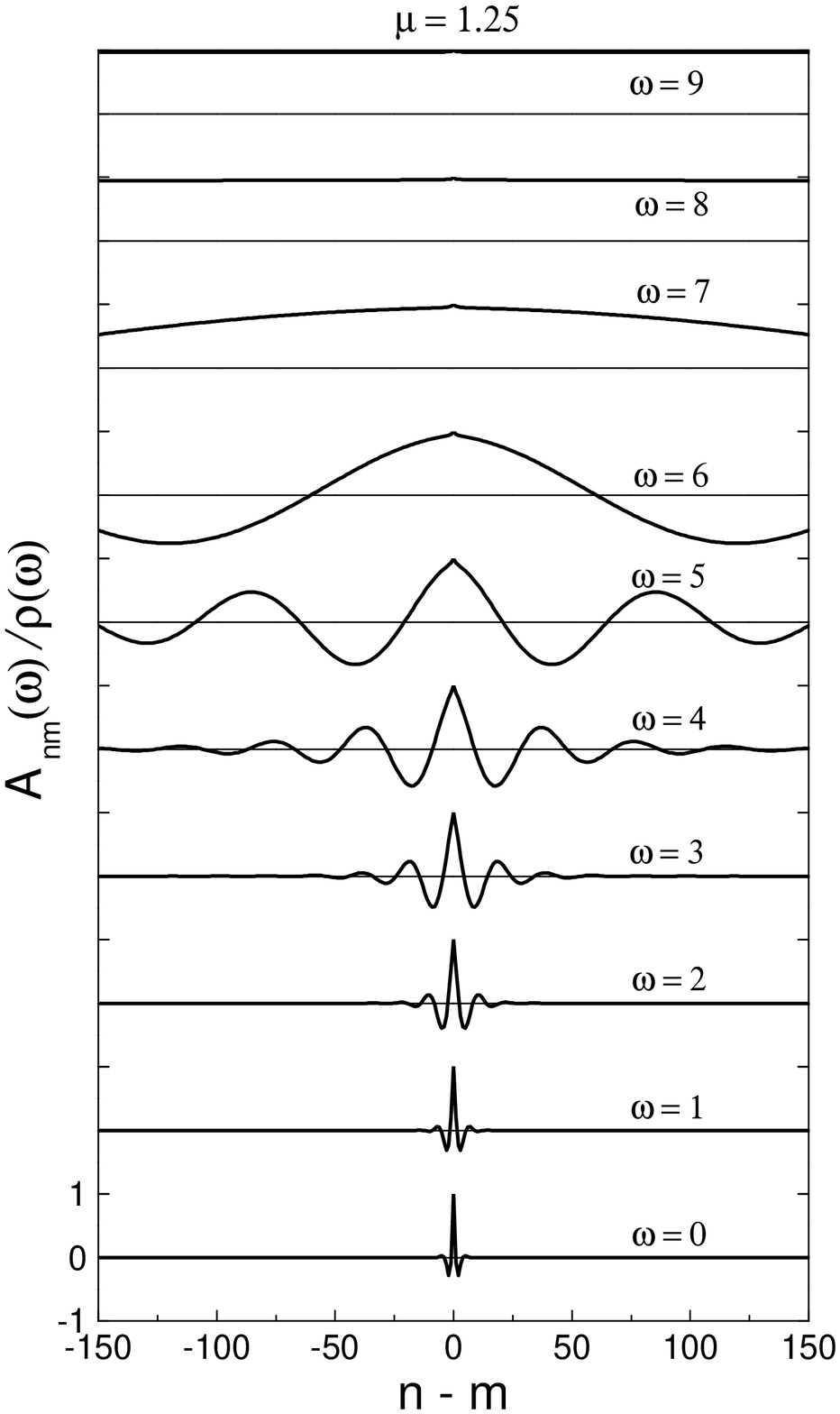}
\includegraphics[height=85mm]{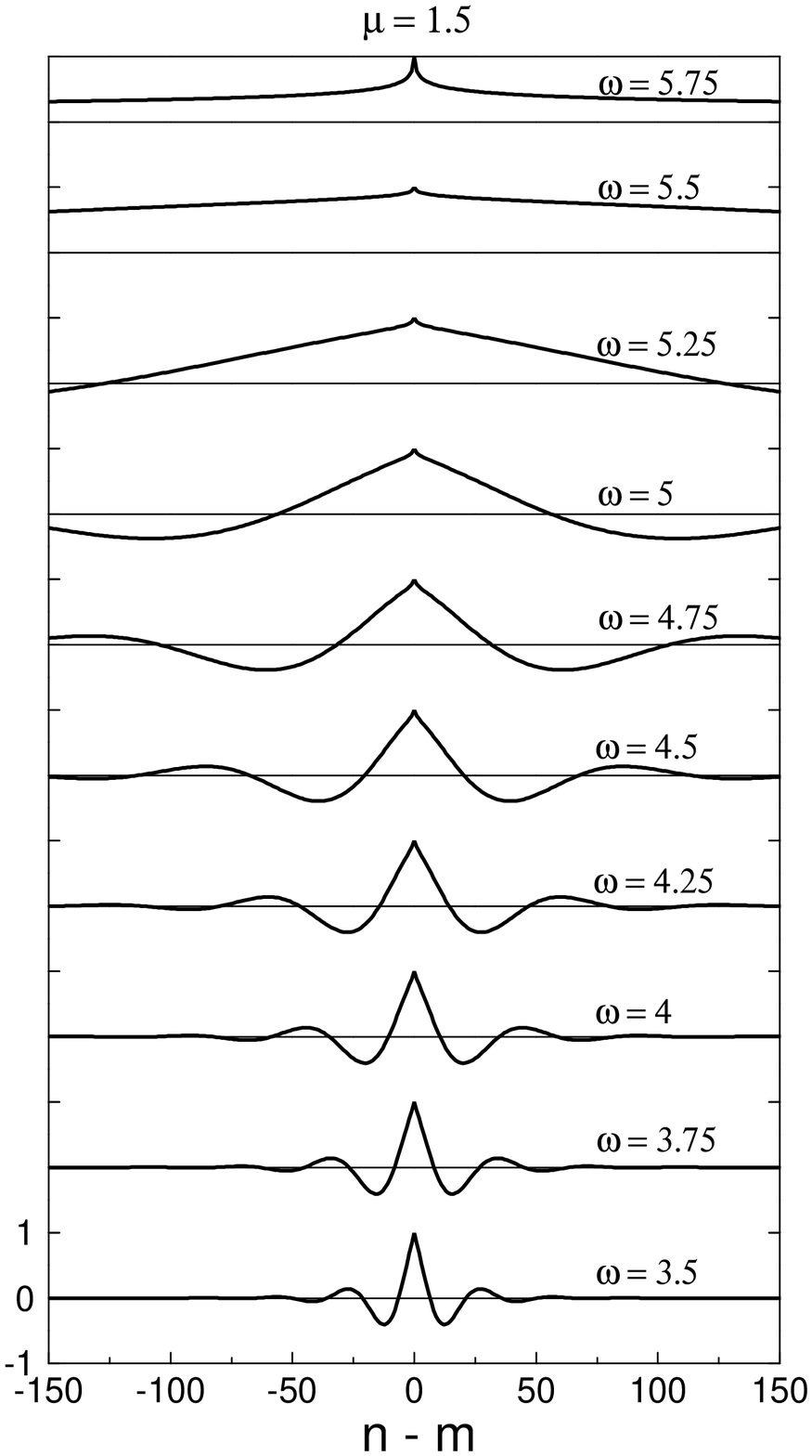}
\includegraphics[height=85mm]{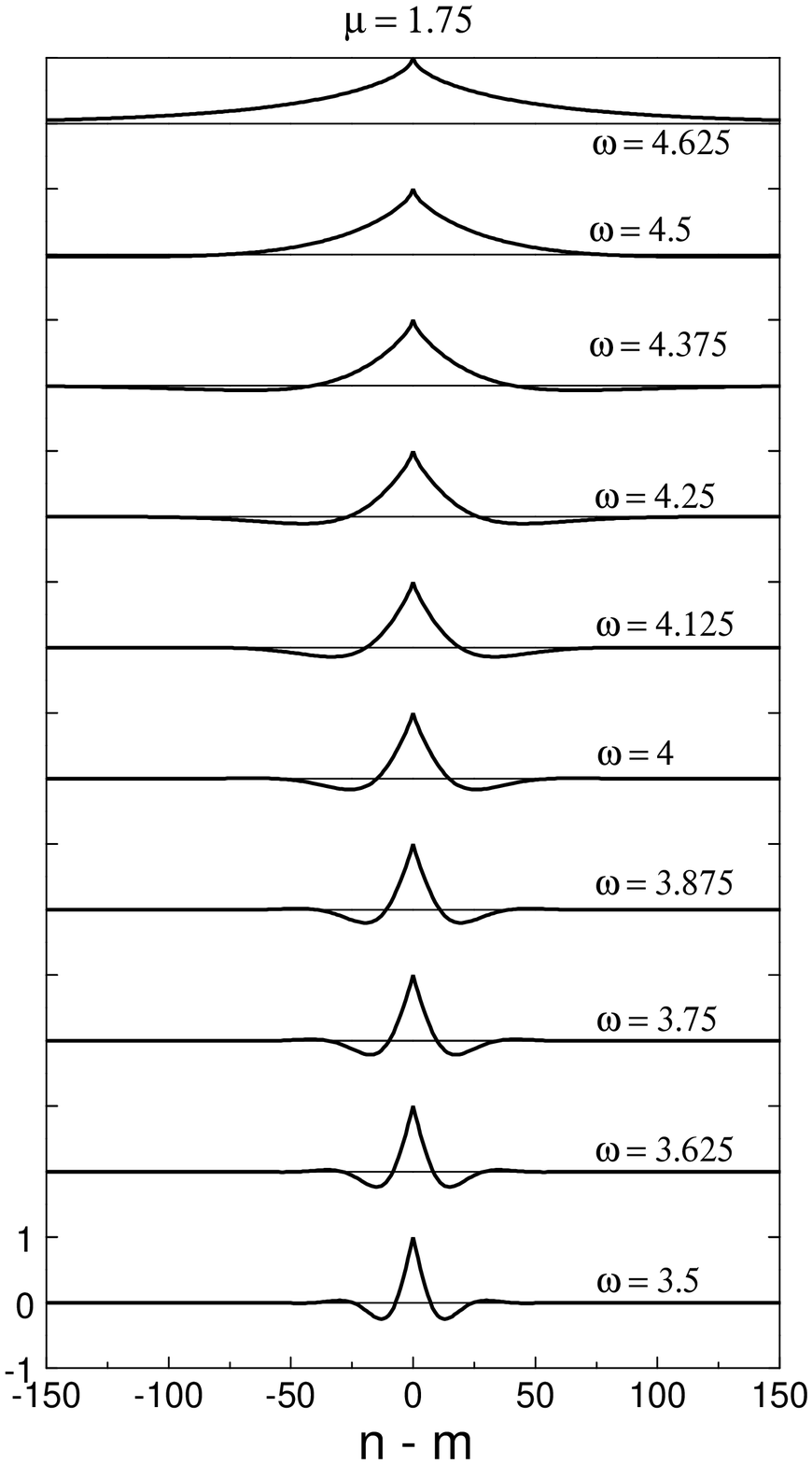}
\caption{Real-space spectral density normalized to the DOS. The
degree of disorder $\Delta$ is $2.5$ for all curves.}
\label{fig:SDReal}
\end{figure*}

So far, we were using the information contained in the
site-diagonal part of the Green's function. Now we turn to the
off-diagonal part of $G(\omega)$ to reveal physics which is
related to the PCL. In Sec.~\ref{sec:PCL} it has been anticipated
that the PCL behaves in an essentially different way in the two
ranges of the hopping exponent $\mu$ separated by $\mu = 3/2$
(marginal case): it remains finite across the band as $\mu
> 3/2$ while having a power-law singularity at the upper band-edge
as $\mu < 3/2$. To illustrate the argumentation used, we plot in
Fig.~\ref{fig:SDMomentum} the momentum-domain spectral density
obtained within the CPA for three values of $\mu$: $\mu < 3/2$,
$\mu=3/2$, and $\mu > 3/2$. The first plot ($\mu = 1.25$) clearly
shows that the width of the resonance along the momentum axis
vanishes as the energy approaches the upper band edge. In
contrast, the same width remains finite for $\mu = 1.75$. The
observed behavior is in complete agreement with the general
definition~(\ref{KappaDef}) and the asymptotic formulas for the
PCL presented in Sec.~\ref{sec:PCL}.

With the same values of $\mu$ as above we also computed the
real-space spectral density $A_{\bf nm}(\omega)$ for several fixed
energies. The quantity plotted in Fig.~\ref{fig:SDReal} is the
real-space spectral density normalized to the DOS: it is unity
when taken about the same site [$\rho(\omega) = A_{\bf
nn}(\omega)$]. For energies located sufficiently far from the
upper band edge, $A_{\bf nm}(\omega)$ displays damped oscillations
as a function of the distance between sites. Their periodicity is
approximately determined by the disorder-free momentum at the
specified energy; naturally, it decreases upon approaching the
upper band edge that in the absence of disorder would correspond
to zero momentum. The inverse damping rate of the oscillations
provides an estimate of the PCL.

\begin{figure}[ht]
\centering
\includegraphics[width=65mm]{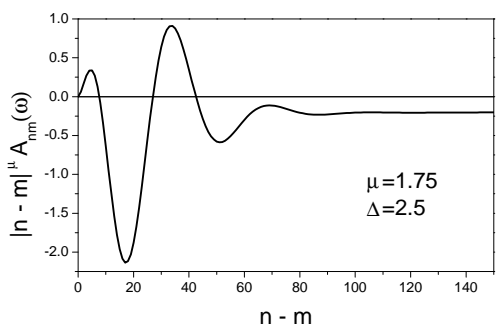}
\caption{Detection of the power-law component in the real-space
spectral density. For the indicated parameters the CPA self-energy
at $\omega = 3.57$ was found to be $\sigma \approx 0.66 - 0.42 i$.
The large-distance saturation value of the plotted quantity
$\approx -0.19$ shows excellent agreement with the analytic result
$(-1/\pi) \,{\rm Im}\, (\omega - \sigma - E_0)^{-2}$.}
\label{fig:SDHop}
\end{figure}

Before discussing PCL in more detail, let us make some remarks on
the spatial behavior of the Green's function. As it was first
mentioned in Sec.~\ref{sec:MomDom}, the real-space Green's
function contains not only a contribution varying exponentially
with coordinate, but also one falling off according to a power
law. In 1D system with arbitrary disorder-induced self-energy, the
main power-law contribution is given by Eq.~(\ref{PowLaw}). The
best way to illustrate the presence of such power-law component is
to plot the product $|{\bf n} - {\bf m}|^\mu A_{\bf nm}(\omega)$
against the coordinate, as is done in Fig.~\ref{fig:SDHop}. The
finite offset of the plotted quantity for large $|{\bf n} - {\bf
m}|$ is in good agreement with that following from the asymptotic
expression~(\ref{PowLaw}).

\begin{figure}[ht]
\centering
\includegraphics[width=70mm]{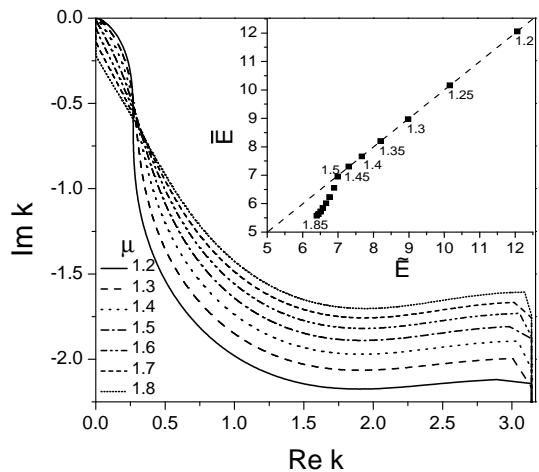}
\caption{Trajectories of the complex roots of the equation $\omega
- \sigma(\omega) = E_k$. For the indicated values of $\mu$, the
critical energy is compared with the upper band edge (see the
inset).} \label{fig:Pole}
\end{figure}

The magnitude of the Green's function component falling off
exponentially with coordinate is determined by the complex roots
$k$ of the equation $E_k = \omega - \sigma(\omega)$. Typical
trajectories of $k(\omega)$ for varying energy are shown in
Fig.~\ref{fig:Pole}. For concreteness, only the solutions with
negative imaginary part were considered. The roots exist only for
$\omega$ below some critical energy $\bar E$. As $\omega$ reaches
$\bar E$, the root disappears encountering the branch cut of the
multivalued function $E_k$ running along the imaginary axis. Hence
the exponential contribution vanishes for all $\omega$ above the
critical energy. In order to elucidate whether the energies
$\omega > \bar E$ still belong to the spectrum we plotted in the
inset to Fig.~\ref{fig:Pole} the critical energy against the upper
band edge for different exponents $\mu$. As follows from this
plot, for $\mu < 3/2$ the energy $\bar E$ coincides with $\tilde
E$, i.e.~the exponentially falling-off component of the Green's
function exists for all $\omega$ up to the band edge. Furthermore,
as can be seen in Fig.~\ref{fig:Pole}, for the above indicated
interval of $\mu$, $\,{\rm Im}\, k(\omega)$ vanishes at $\omega =
\tilde E$. Hence, the exponential component of the Green's
function will have constant envelope at the upper band-edge,
thereby extending infinitely in space. This behavior is indeed
observed in the panel of Fig.~\ref{fig:SDReal} with $\mu=1.25$,
thus confirming the main statement of the paper about the
divergence of the PCL. Regarding the interval $\mu > 3/2$, it
follows from the inset of Fig.~\ref{fig:Pole} that the energy
$\bar E$ lies inside the band. In this case, the PCL remains
finite across the band as is illustrated in Fig.~\ref{fig:SDReal}
(panel of $\mu = 1.75$).

\section{Summary and concluding remarks \label{sec:Summary}}

In summary, we have investigated a tight-binding model with
uncorrelated diagonal disorder and nonrandom hopping integrals
given by $J_{\bf nm} = 1/|{\bf n -m}|^\mu$. Using the coherent
potential approximation, we calculated the density of states and
the phase coherence length related to this system. The addressed
quantities were found to be affected by disorder in an essentially
different way depending on whether the hopping range exponent lies
in the interval $d < \mu < 3d/2$ or $\mu > 3d/2$. The first of
these intervals is featured by the fact that, for turned-off
disorder, the infinite slope of the dispersion relation produces a
tail of the density of states at the high-energy part of the
spectrum. The effect of disorder, which introduces the mixing of
states only within a narrow energy interval, is seriously weakened
due to the small number of available states. In accordance with
such qualitative argument, the tail-like part of the density of
states, found in a self-consistent way, remains similar to that of
the disorder-free system, apart for a small uniform shift induced
by disorder. For $\mu < 3d/2$ the zero broadening at the edge of
the tail was shown to result in the divergence of the phase
coherence length, while for $\mu > 3d/2$ it remained finite across
the band. Irrespectively of $\mu$, the real-space propagator was
demonstrated to contain a component decaying with coordinate as
$1/|{\bf n-m}|^\mu$.

The model studied in the present paper is applicable to various
materials in which the energy of one-particle excitations is of
dipolar origin. As an example, let us mention dipolar Frenkel
excitons on two-dimensional regular lattices where molecules are
subjected to randomness due to a disordered
environment.~\cite{Nabetani95} Some biological light-harvesting
antenna systems~\cite{vanAmerongen00} as well as
dendrimers~\cite{Kopelman97} may represent a realization of this
model. The form of dipolar interactions dictates the value of the
hopping exponent, $\mu=3$. For this value of $\mu$, $d=2$ was
shown to be the marginal case at which one starts to observe the
divergence of the PCL at the upper band-edge. Increasing the PCL
for Frenkel excitons could be deduced after measurement of the
linear absorption spectra, that in this case should fit the
power-law form~(\ref{Absorption}). This opens the feasibility to
test our predictions from experiments at low temperature.

\begin{acknowledgments}

We are grateful to G.\ C.\ La Rocca and V.\ M.\ Agranovich for
helpful discussions. D.\ B.\ B.\ thanks Universidad Complutense
for hospitality. V.\ A.\ M.\ acknowledges support from NATO during
the initial stage of this work. Work in Pisa was supported by MIUR
(PRIN-2001). Work in Madrid was supported by DGI-MCyT
(MAT2003-01533).

\end{acknowledgments}

\appendix

\section{Evaluation of the real-space propagator (1D system) \label{app:oned}}

\begin{figure}[bt]
\centering
\includegraphics[height=30mm]{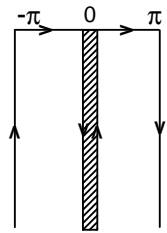}
\caption{Complex-momentum integration contour for $n-m < 0$. The
contour corresponding to $n-m > 0$ is obtained by mirror
reflection with respect to the horizontal axis.}
\label{fig:Contour}
\end{figure}

In this appendix, we calculate the real-space propagator for $d=1$
to show that the presence of LRH makes it vary in space
nonexponentially. Following Ref.~\onlinecite{Balagurov}, we
compute
\begin{equation}
\label{GreenN}
G_{nm}(\omega) = \int_{-\pi}^{\pi} \frac{dk}{2\pi}
\, \frac{e^{ik(n-m)}}{\omega - \sigma - E_k}
\end{equation}
by extending the integration contour in the complex momentum plane
as illustrated in Fig.~\ref{fig:Contour}. The $n\,$th term in
Eq.~(\ref{EnergyK}), decreasing not rapidly enough for $n \to
\infty$, renders $E_k$ a multivalued function of the complex $k$:
for all integer $s$, branch cuts appear connecting points $2\pi s$
with $2\pi s \pm i\infty$. Integral~(\ref{GreenN}) can be split
into a part due to poles and one due to the branch cuts of the
integrand
\begin{align}
\label{Prop2}
G_{nm}(\omega) & = -i \left( \frac{dE_k}{dk}\right)^{-1}_{k(\omega)}
e^{-i k(\omega) |n-m|} \nonumber \\
& -i\int_0^{+\infty}\frac{d\kappa}{2\pi}\,e^{-\kappa|n-m|} \nonumber \\
& \times  \left[ \frac{1}{\omega - \sigma - E_{i\kappa-0}}
- \frac{1}{\omega - \sigma - E_{i\kappa+0}} \right]\ .
\end{align}
The first term, in which pole $k(\omega)$ is to be obtained from
equation $\omega - \sigma(\omega) = E_k$ ($|\, {\rm Re} \, k| <
\pi$, $\,{\rm Im}\, k < 0$), falls off as an exponential of
coordinate. This behavior is similar to that observed for a full
real-space propagator in the case of SRH. We show now that in the
presence of LRH, the second term decays as a power of distance. In
the limit of $|n - m| \to +\infty$ the upper cutoff of the
integral entering Eq.~(\ref{Prop2}) can be replaced with $1/|n-m|
\to 0$. Hence, only few terms in the expansion of $E_{i \kappa \pm
0}$ around $\kappa = 0$ are sufficient to know large-$|n-m|$
asymptotic. The 1D version of Eq.~(\ref{ExpansionGeneral}) reads
$E_k = 2 \zeta(\mu) + \Gamma(1-\mu) [(ik)^{\mu - 1} +
(-ik)^{\mu-1}]$, where $\zeta$ and $\Gamma$ are the Riemann Zeta
function and Gamma function, respectively.~\cite{Wolfram} After
some algebra we get
\begin{equation}
\label{PowLaw} G_{nm}(\omega) \sim \frac1{[\omega - \sigma(\omega)
- E_0]^2} \, \frac1{|n-m|^\mu}\ ,
\end{equation}
where $E_0 = 2\zeta(\mu)$ is the disorder-free upper band-edge
energy.


\begin{thebibliography}{99}

\bibitem{Anderson} P.\ W.\ Anderson, Phys.\ Rev.\ \textbf{109}, 1492 (1958).

\bibitem{Gang4} E.\ Abrahams, P.W.\ Anderson, D.C.\ Licciardello, and V.\
    Ramakrishnan, Phys.\ Rev.\ Lett.\ \textbf{42}, 673 (1979).

\bibitem{LocalizedOneD} N.\ F.\ Mott and W.\ D.\ Twose, Adv.\ Phys.\
    \textbf{10},  107 (1961); B.\ Kramer and A.\ MacKinnon, Rep.\ Prog.\
    Phys.\ \textbf{56}, 1469 (1993).

\bibitem{Cressoni98} J.\ C.\ Cressoni and M.\ L.\ Lyra, Physica A \textbf{256},
    18 (1998).

\bibitem{Rodriguez00} A.\ Rodr\'{\i}guez, V.\ A.\ Malyshev, and F.\
        Dom\'{\i}nguez-Adame, J.\ Phys.\ A: Math.\ Gen.\ {\bf 33}, L161
    (2000).

\bibitem{MalyshevPRL} A.\ Rodr\'{\i}guez, V.\ A.\ Malyshev, G.\ Sierra, M.\ A.\
    Mart\'{\i}n-Delgado, J. Rodr\'{\i}guez-Laguna, and F.\
    Dom\'{\i}nguez-Adame, Phys.\ Rev.\ Lett.\ \textbf{90}, 27404 (2003).

\bibitem{Xiong03} S.-J.\ Xiong and G.-P.\ Zhang, Phys.\ Rev. B
    \textbf{68}, 174201 (2003).

\bibitem{PLLoc} C.\ Yeung and Y.\ Oono, Europhys.\ Lett.\ \textbf{4}, 1061
    (1987).

\bibitem{CPARef} P.\ Soven, Phys.\ Rev.\ \textbf{156}, 809 (1967); D.V.\
    Taylor, Phys. Rev.\ \textbf{156}, 1017 (1967).

\bibitem{EliottRev} R.\ J.\ Elliott, J.\ A.\ Krumhansl, P.\ L.\ Leath, Rev.\
    Mod.\ Phys.\ \textbf{46}, 465 (1974).

\bibitem{LitScaling} A.\ Boukahil and D.\ L.\ Huber, J. Lumin. {\textbf 45},
        13 (1990); M.\ Schreiber and Y. Toyozawa, J.\ Phys.\ Soc.\ Jpn.\
        \textbf{51}, 1528 (1982); H.\ Fidder, J.\ Knoester and D.\ A.\ Wiersma,
        J.\ Chem.\ Phys.\ \textbf{95}, 7880 (1991); L.\ D.\ Bakalis and J.
        Knoester, J. Lumin. {\textbf 86-87}, 66 (2000).

\bibitem{Malyshev91} V.\ A.\ Malyshev Opt.\ Spektrosk.\ {\textbf 71}, 873
        (1991) [Opt.\ Spectrosc. {\textbf 71}, 505 (1991)]; J.\ Lumin.
        {\textbf 55}, 225 (1993); V.\ Malyshev and P.\ Moreno, Phys.\ Rev.\ B
        {\textbf 51} 14587 (1995); V.\ A.\ Malyshev, A.\ Rodr\'{\i}guez,
        and F.\ Dom\'{\i}nguez-Adame, Phys. Rev. B {\bf 60},
        14140 (1999); V.\ A.\ Malyshev and F.\ Dom\'{\i}nguez-Adame, Chem.\
        Phys. Lett. {\bf 313}, 255 (1999); A.\ V.\ Malyshev  and V.\ A.\
        Malyshev, Phys. Rev. B {\bf 63}, 195111 (2001).

\bibitem{Knapp84} E.\ W.\ Knapp, Chem.\ Phys.\ \textbf{85}, 73 (1984).

\bibitem{Abrikosov} A.\ A.\ Abrikosov, L.\ P.\ Gorkov, and I.\ E.\ Dzyaloshinski,
    \emph{Methods of Quantum Field Theory in Statistical Physics\/} (Dover,
    New York, 1975)

\bibitem{Moments} B.\ Velicky, S.\ Kirkpatrick, and H.\ Ehrenreich, Phys.\ Rev.
    \textbf{175}, 747 (1968).

\bibitem{Balagurov} D.\ B.\ Balagurov, G.\ C.\ La Rocca, and V.\ M.\
    Agranovich, Phys.\ Rev.\ B \textbf{68}, 45418 (2003).

\bibitem{Wolfram} To derive this formula one can notice that $E_k =
    \mbox{Li}_\mu(e^{ik}) + \mbox{Li}_\mu(e^{-ik})$, where $\mbox{Li}$ is
    Polylogarithm, and then use the expansion presented at

        {\tt http://functions.wolfram.com/10.08.06.0024.01}

\bibitem{Elliott} S.\ N.\ Taraskin, Y.\ L.\ Loh, G.\ Natarajan, and S.\ R.\
    Elliott, Phys.\ Rev.\ Lett.\ \textbf{86}, 1255 (2001); J.\ J.\ Ludlam,
    S.\ N.\ Taraskin, and S.\ R.\ Elliott, Phys.\ Rev.\ B \textbf{67},
    132203 (2003).

\bibitem{Nabetani95} A.\ Nabetani, A.\ Tamioka, H.\ Tamaru, and K.\ Miyano,
        J.\ Chem.\ Phys.\ {\bf 102}, 5109 (1995); A.\ Tamioka and K.\ Miyano,
        Phys.\ Rev.\ B {\bf 54}, 2963 (1996); F.\ Dom\'{\i}nguez-Adame,
        V.\ A.\ Malyshev, and A.\ Rodr\'{\i}guez, J.\ Chem.\ Phys.\ {\bf 112},
        3023 (2000); L.\ D.\ Bakalis, I.\ Rubtsov, and J.\ Knoester, J.\ Chem.\
        Phys. {\bf 117}, 5393 (2002); S.\ S.\ Lampoura, C.\ Spitz, S.\ Dahne,
        J.\ Knoester, and K.\ Duppen, J.\ Phys.\ Chem.\ B {\bf 106}, 3103 (2002).

\bibitem{vanAmerongen00}H.\ van Amerongen, L.\ Valkunas, and R.\ van Grondelle,
        {\it Photosynthetic Excitons}, World Scientific: Singapore, 2000;
        T.\ Renger, V.\ May, and O.\ K\"uhn, Phys. Rep. {\bf 343}, 137 (2001).

\bibitem{Kopelman97} R.\ Kopelman, M.\ Shortreed, Z.-Y.\ Shi, W.\ Tan, Z.\ Xu,
        J.\ Moore, A.\ Bar-Haim, and J.\ Klafter, Phys.\ Rev.\ Lett.\ {\bf 78},
        1239 (1997); K.\ Herigaya, Phys.\ Chem.\ Chem.\ Phys.\ {\bf 1},
        1687 (1999);  M.\ A.\ Mart\'{\i}n-Delgado, J.\ Rodr\'{\i}guez-Laguna,
        G.\ Sierra, Phys.\ Rev.\ B {\bf 65}, 155116 (2002); O.\ P.\ Varnavski,
        J.\ C.\ Ostrowski, L.\ Sukhomlinova, R.\ J.\ Twieg, G. C.\ Bazan,
        and T.\ Goodson III, J. Am. Chem. Soc. {\bf 124}, 1736 (2002).

\end{thebibliography}
\end{document}